\documentclass[twocolumn,tighten]{aastex63}

\pdfoutput=1 
\usepackage{amsmath,amstext,longtable}
\usepackage[T1]{fontenc}
\usepackage{apjfonts} 
\usepackage[figure,figure*]{hypcap}
\usepackage{hyperref}
\usepackage{tablefootnote}
\received{TBD}
\revised{TBD}
\accepted{TBD}
\submitjournal{AJ}
\shortauthors{Kota et al.}
\shorttitle{Nearby Discoveries}

\usepackage{color,amsmath,longtable}

\begin{document}

\title{Discovery of 16 New Members of the Solar Neighborhood using Proper Motions from CatWISE2020}

\correspondingauthor{Tarun Kota}
\email{Tkota0910@gmail.com}

\author[0000-0001-8764-7648]{Tarun Kota}
\affiliation{Eastview High School, 6200 140th St W, Apple Valley, MN 55124, USA;}

\author[0000-0003-4269-260X]{J.\ Davy Kirkpatrick}
\affiliation{IPAC, Mail Code 100-22, Caltech, 1200 E. California Blvd., Pasadena, CA 91125, USA}

\author[0000-0001-7896-5791]{Dan Caselden}
\affiliation{Department of Astrophysics, American Museum of Natural History, Central Park West at 79th Street, New York, NY 10034, USA}

\author[0000-0001-7519-1700]{Federico Marocco}
\affiliation{IPAC, Mail Code 100-22, Caltech, 1200 E. California Blvd., Pasadena, CA 91125, USA}

\author[0000-0002-6294-5937]{Adam C.\ Schneider}
\affiliation{United States Naval Observatory, Flagstaff Station, 10391 West Naval Observatory Rd., Flagstaff, AZ.86005, USA}
\affiliation{Department of Physics and Astronomy, George Mason University, MS3F3, 4400 University Drive, Fairfax, VA 22030, USA}

\author[0000-0002-2592-9612]{Jonathan Gagn\'e}
\affiliation{Plan\'etarium Rio Tinto Alcan, Espace pour la Vie, 4801 av. Pierre-de Coubertin, Montr\'eal, Qu\'ebec, Canada}
\affiliation{Institute for Research on Exoplanets, Universit\'e de Montr\'eal, D\'epartement de Physique, C.P. 6128 Succ. Centre-ville, Montr\'eal, QC H3C 3J7, Canada}

\author[0000-0001-6251-0573]{Jacqueline K.\ Faherty}
\affiliation{Department of Astrophysics, American Museum of Natural History, Central Park West at 79th Street, New York, NY 10034, USA}

\author[0000-0002-1125-7384]{Aaron M.\ Meisner}
\affiliation{NSF's National Optical-Infrared Astronomy Research Laboratory, 950 N. Cherry Avenue, Tucson, AZ 85719, USA}

\author[0000-0002-2387-5489]{Marc J.\ Kuchner}
\affiliation{NASA Goddard Space Flight Center, Exoplanets and Stellar Astrophysics Laboratory, Code 667, Greenbelt, MD 20771, USA}

\author[0000-0003-2478-0120]{Sarah Casewell}
\affiliation{School of Physics and Astronomy, University of Leicester, University Road, Leicester LE1 7RH, UK}

\author[0000-0001-9878-0436]{Kanishk Kacholia}
\affiliation{Department of Computer Science, University of Minnesota, 200 Union St SE, Minneapolis, MN 55455, USA}

\author[0000-0003-2235-761X]{Tom Bickle}
\affiliation{Backyard Worlds}

\author{Paul Beaulieu}
\affiliation{Backyard Worlds}

\author[0000-0002-7630-1243]{Guillaume Colin}
\affiliation{Backyard Worlds}

\author[0000-0002-7389-2092]{Leslie K.\ Hamlet}
\affiliation{Backyard Worlds}

\author[0000-0002-7587-7195]{J\"org Sch\"umann}
\affiliation{Backyard Worlds}

\author{Christopher Tanner}
\affiliation{Backyard Worlds}

\author{The Backyard Worlds: Planet 9 Collaboration}

\begin{abstract}
\par
In an effort to identify nearby and unusual cold objects in the solar neighborhood, we searched for previously unidentified moving objects using CatWISE2020 proper motion data combined with machine learning methods. We paired the motion candidates with their counterparts in 2MASS, UHS, and VHS. Then we searched for white dwarf, brown dwarf, and subdwarf outliers on the resulting color-color diagrams. This resulted in the discovery of 16 new dwarfs including two nearby M dwarfs (< 30 pc), a possible young L dwarf, a high motion early T dwarf and 3 later T dwarfs. This research represents a step forward in completing the census of the Sun's neighbors.
\end{abstract}


\section{Introduction}
 The Wide-Field Infrared Survey Explorer (\textit{WISE}; \citealt{Wright 2010}), with its unique full-sky sensitivity, has the potential to answer fundamental astrophysical questions. From 2009-2011, \textit{WISE} gave astronomers unparalleled photometric data on objects across the span of the universe. Using these data, there were many \textit{WISE}- based discoveries including the first Earth Trojan Asteroid (\citealt{Connors 2011}), the most luminous galaxy (\citealt{Tsai 2015}), and the coldest and closest known brown dwarfs (\citealt{Luhman 2013}, \citealt{Luhman 2014}). Due to cryogen exhaustion in late 2010, \textit{WISE}  was forced to change from a four-band Survey (W1 (3.4 $\mu$m), W2 (4.6$\mu$m), W3 (12$\mu$m), W4 (22$\mu$m)) into a two band survey (W1, W2). The mission was repurposed as NEOWISE (\citealt{Mainzer 2011}). Since its launch, NEOWISE has detected over 34,000 minor planets (\citealt{Mainzer 2014}), as near-earth object detection is the new scientific driver of the mission.	
\par
Nevertheless, for astronomers who were interested in identifying motion objects outside of our solar system, the only available \textit{WISE} processing that leveraged the time series data was the AllWISE catalog (\citealt{Cutri 2013}). However, AllWISE only uses early \text{WISE} data (13 months viewing time), which means it is only able to identify brighter, high motion objects (\citealt{Kirkpatrick 2014}, \citealt{Kirkpatrick 2016}). With a short time baseline, our ability to detect fainter moving objects and smaller motion is limited. The key factor is that fainter objects have larger positional uncertainties, which prohibits us from finding faint, smaller motion objects.

 The CatWISE Preliminary Catalog (\citealt{Eisenhardt 2020}) is a reprocessing of \textit{WISE}/NEOWISE data over a 6 year time frame (2010-2016) to measure motions of more slowly moving objects. Since NEOWISE is only a W1 and W2 survey, CatWISE Preliminary only has these two filters. CatWISE Preliminary - the richer data set - enabled motion detection at fainter magnitudes and smaller motions.
\par
CatWISE2020 (\citealt{Marocco 2021}) took this one step further by adding another two years of data to that used for the CatWISE Preliminary Catalog, bringing the total to six times as many exposures spanning over 16 times as large a time baseline as the AllWISE catalog. In addition, the detection list for the CatWISE2020 Catalog was generated using the unWISE Catalog (\citealt{Schlafly 2019}) instead of the AllWISE detection software used by the CatWISE Preliminary pipeline. These two factors led to the CatWISE2020 Catalog having almost twice as many sources as the CatWISE Preliminary Catalog. Although this improved astrometric accuracy has led to initial discoveries (\citealt{Kirkpatrick 2021}, \citealt{Meisner 2021},  \citealt{Rothermich 2021}), the CatWISE2020 data are still largely unexplored.
\par
In this paper, we use the CatWISE2020 proper motions to search for nearby and unusual cold objects that have otherwise been overlooked. Specifically, due to their low luminosity, cooler brown dwarfs comprise the majority of gaps in our knowledge of the solar neighborhood (\citealt{Kirkpatrick 2019}). However, M dwarfs, which are the most common objects in the solar neighborhood, can also be missing due to small motions because earlier surveys may have lacked sufficient time baseline to identify them. 

A complete census of astronomical objects in the solar neighborhood is important in determining the low mass cutoff for star formation (\citealt{Kirkpatrick 2019}) and would allow researchers to do more targeted searches for habitable exoplanets around these host objects (\citealt{Gillon 2017}, \citealt{Zechmeister 2009}).
\par
In this paper, we present the discovery of 16 new objects including two nearby M dwarfs, a likely young L dwarf, a high motion early T dwarf and 3 additional late T dwarfs. In \S~\ref{Candidate Selection} we discuss the method used to select candidates from CatWISE2020. In \S~\ref{Analyzing Candidates} we discuss the identification methods used to select interesting objects from our candidate list. In \S~\ref{Interesting objects} we characterize our most interesting new discoveries. In \S~\ref{Conclusion} we give brief conclusions.


\section{Candidate Selection\label{Candidate Selection}}
We used XGBoost (\citealt{Chen 2016}) to identify rows in CatWISE2020 that may pertain to moving objects. XGBoost is a feature-rich machine learning package centered around gradient boosting. From XGBoost, we chose to train binary classification trees that output a single floating point score with values from 0.0 to 1.0 for each CatWISE2020 row. Lower scores indicate the row is less like those of moving objects in our training set, whereas higher scores mean the row is more like them. Typically, we would explore classifier results by sorting them with higher scores first, and stop exploring results when we encounter so many false positives that we are no longer effectively using our time. Although every classifier we trained was different, and had completely different score distributions, we were usually overwhelmed by false positives well before scores of 0.5.

\par
Being a binary classification task means that the classifiers are trained to answer a “yes” or “no” question, or, thought of another way, predict membership of two classes. For our classifiers, these are the “moving object” class and the “non-moving object” class. The moving object class asks “does the object to which this CatWISE2020 row pertains have significant proper motion?” Significant proper motion, for our methodology, is a proper motion that a human verifier could confirm by watching an animation of WISE imagery spanning approximately 10 years. This ends up being around 50 to 150 mas/yr, but is different for every human verifier and candidate moving object’s signal-to-noise ratio. The non-moving object class is then the complement of the moving object class: if the CatWISE2020 row does not pertain to an object with significant proper motion, it is a member of the non-moving object class.
\begin{figure*}
\figurenum{1}
\includegraphics[scale=0.5,angle=0]{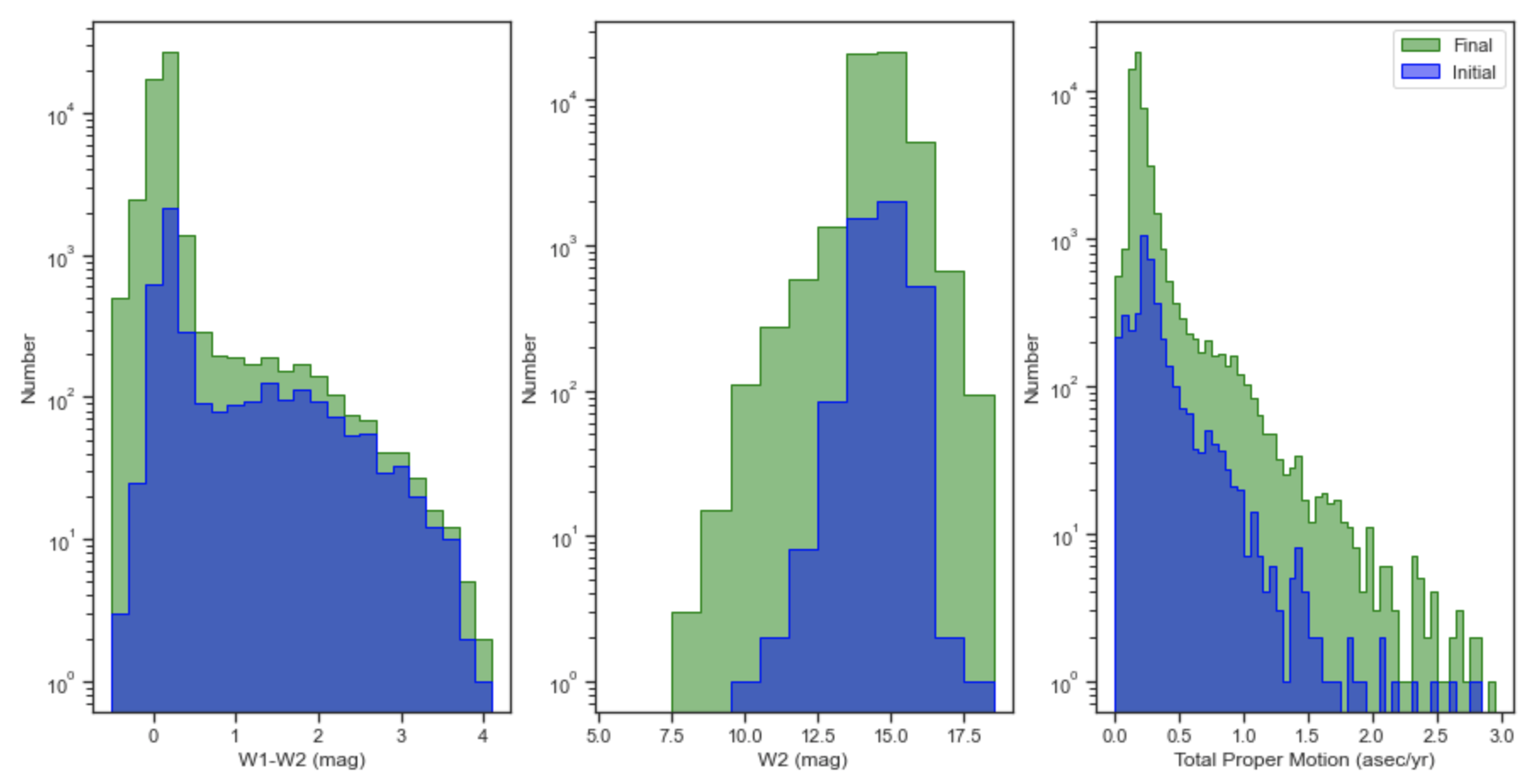}
\centering
\caption{ These histograms show the distribution of the initial (blue) and final (green) moving object training sets across W1-W2, W2 and total proper motion. Note for our W1-W2 histogram we only plotted objects that had W1-W2 values between $-$0.5 and 4.0 mag.
\label{histogram}}
\end{figure*}
\par
\par
To construct our dataset, we took a random sample of about 200,000 rows from CatWISE2020 and temporarily assumed that none of them pertained to moving objects. This was our non-moving object training set. The classifier would train to output scores that are low for rows like these.
\par
Next, we took the results of a previous candidate search that contained coordinates for candidate moving objects. These objects had previously been human reviewed, and in the human’s opinion, they exhibited discernable proper motion in WISE animations. We cross-matched these candidate moving objects with CatWISE2020 to obtain CatWISE2020 rows for each candidate. Some candidates did not have matches, some matched incorrect rows, and some matched more than one correct row. More than one correct row can happen when CatWISE2020 fits multiple detections of the same source. We plotted the cross-match results in various views such as with color, magnitude, and cross-match angular separation to identify obvious spurious matches by their outlying from the majority of the dataset, and eliminated these matches from our training set. We then temporarily assumed that all of the remaining matches were correct. This left us with about 4000 rows. This was our moving object training set.
\par
We next assigned class weights to the moving and non-moving object training set. With an imbalanced training set, weighting is crucial.
As XGBoost builds classifiers, it runs partially built classifiers against the training set and measures the difference between the classifiers' predicted versus desired output. That difference goes into the loss function (we used the default log loss function), which is then used to calculate loss. Without any weighting, the non-moving set will contribute $\sim$200,000 / $\sim$4000 $\approx$ 50 times more examples to the loss function than the moving set.
As a result, the classifier will be incentivized to return a low score and will not train as well. To address this, class weighting is the standard approach. We chose a “balanced” strategy that equalizes the total weight of all members of each class. Members of the overpopulated non-moving set get a lower weight, and the training loss is scaled down by that lower weight. Members of the underpopulated moving set get a higher weight, and the training loss is scaled up.
\par
In addition to weighting by class, we also weighted the moving set using the PSF-fit magnitude in W2 (w2mpro in the CatWISE2020 Catalog). We divided the moving set into 0.5-mag bins, and equalized the total weight in each bin. In \cite{Marocco 2019}, balancing weights by w2mpro was motivated by the recognition that our moving object set contained few very faint examples, and we desired to improve the model’s performance on very faint examples. We also limited rows to w2mpro $>$ 14.0 mag under the assumption that bright objects had been exhausted by prior searches and we hoped that the reduced scope would make the model perform better for faint objects. These processes are described in more detail in \S3 of \cite{Marocco 2019}.
\par
We then removed columns that were inappropriate for classification, such as coordinates and source identifiers. Then, we added new columns, commonly called features, calculated from catalog parameters. For example, we added signal to noise of the W1 versus W2 profile-fit photometry $w1snr - w2snr$; signal to noise of profile-fit photometry without versus with the CatWISE2020 astrometric solution such as $w1snr - w1snr\_pm$; and differences in aperture photometry in the same band but with another aperture size such as 5.5" versus 8.25" $w1mag\_1 - w1mag\_2$ and 8.25" versus 11" $w1mag\_2 - w1mag\_3$. We exhaustively added, removed, and modified features between training countless classifiers until improvements to the classifier’s performance were no longer statistically significant across multiple training runs. We are under no illusion that our feature set is optimal. Instead, we experimentally found that removing any individual features from the list did not improve performance, and exhausted our ideas for new features.
\par
Before training the classifier, we removed duplicate rows, randomly shuffled rows, calculated additional feature columns, and dropped non-feature columns from the training set. Then we isolated 20\% of the training set and set it aside. This would be the validation set, used to estimate the classifier’s performance during training. When the classifier’s performance on the validation set stopped improving, the training process would halt. Halting training early in this way is a common technique called “early stopping”. Since each training iteration in XGBoost adds complexity to the classifier, early stopping prevents classifiers from becoming unnecessarily complex.
\par
Then we ran the trained classifier over all CatWISE2020 rows having w2mpro $>$ 14.0 mag. This involves, for each row, calculating additional feature columns, dropping non-feature columns, and passing it to the classifier’s prediction function which returns the score. We performed this operation on all CatWISE2020 rows within one unWISE tile footprint, or about 1/17000th of the sky, per invocation to realize performance gains. For each row with a classifier score over 0.5, we added the score to the original CatWISE2020 row and saved the row to disk.
\par
We then sorted the resulting rows by score, with the highest score first, and examined animations of WISE images at the coordinates for each row. We used WiseView (\citealt{Caselden 2018}), which is an interactive online image blinking tool designed explicitly for human verification of motion candidates in WISE imagery. When the row pertained to a moving object, we would record the row as a true positive. When unsure, we recorded the row as unknown. After examining candidates sorted simply by score, we
various strategies, such as restricting rows by W1-W2 color, proper motion, W2 magnitude, position in the sky, or a combination of these constraints.
\par
Eventually, our sorting returned so few true positives per false positive, and we stopped reviewing results. We then added all of the true positives to the moving object class of the training set, removed rows that we had marked as unknown from the training set entirely, and added all remaining rows to the non-moving object class of the training set. Adding true positives to the moving objects class gave future classifiers more examples, thus making the classifiers more accurate. By adding false positives to the non-moving object class, we add examples that are manually vetted and represent particularly hard examples that challenged our classifier. Both properties make these examples valuable additions to the otherwise randomly sampled non-moving class.
\par
We repeated this process many times: training a classifier, vetting the results, and expanding the training set. Periodically, we purified our training set by running a classifier over the training set itself and re-inspecting highly scored members of the non-moving class and lowly scored members of the moving class. Doing so identified erroneous entries that arrived in our training set through the initial random sample that created the non-moving class, the cross-match that created the moving class, and human process errors such as copy-pasting coordinates into the wrong list.
\par
After some iterations, new classifiers began scoring proper motion stars higher than their predecessors. Our initial moving object training set consisted more of fainter and higher proper motion objects, like brown and white dwarfs, than lower motion main sequence stars. As we added more moving objects to our training set, they inevitably included more objects outside of the brown dwarf and white dwarf set. The classifiers we trained on the updated training set then became more sensitive to proper motion stars of all kinds. This created a feedback loop, the effects of which are apparent in Figure \ref{histogram} that shows distributions of the initial and final moving object training sets. The initial training set is redder in W1 - W2 colors than the final training set. This is an effect of updating the training set for each successive classifier and accepting main sequence stars.
\par
Proper motion main sequence stars—despite overwhelming the human verifier—are not true contaminants in our search. Since they are so well recovered by Gaia and so plentiful, careful accounting was not warranted. To save time for the human verifier, we cross-matched all CatWISE2020 rows that scored higher than 0.5 against a subset of Gaia DR1 with total proper motion $>$ 100 mas/yr and a separation of $<$ 6 arcseconds. The proper motion and angular separation constraints were added to reduce false matches. Over  results from one classifier, this process automatically labeled approximately 15000 rows. After incorporating these automatic labels into our training set and training another classifier, this process automatically labelled approximately 24000 rows. On the third repetition, we changed our cross-match criteria to require total proper motion $>$ 150 mas/yr and separation $<$ 5 arcseconds. The third repetition resulted in about 13000 automatic labels. These automatic labels, like all manual labels, were subjected to our periodic purification processes that aimed to eliminate bad matches.
\par
In early iterations, each new classifier raised many previously unseen moving objects. With each iteration, however, the number of new moving objects decreased. We continued iterating until, ultimately, there were too many false positives per true positive to continue. Once the yield of new sources slowed significantly, we halted the training of other classifiers with this methodology and stopped verifying results from our existing classifiers. Of the motion candidates selected by our classifiers, we eliminated those for which there existed a Gaia DR1 astrometric solution. This left us with $\sim$6000 candidates, both known and unknown, for further analysis.

\section{Analyzing Candidates\label{Analyzing Candidates}}
\subsection{Other Databases}
\par 
As stated earlier, CatWISE2020 only contains two magnitudes (W1 and W2). This meant that we only had access to one color (W1$-$W2) when analyzing our candidates. With only one color, it is difficult to characterize objects. Thus, in order to get more colors, we gathered additional infrared data from Two Micron All-Sky Survey (2MASS: \citealt{Skrutskie 2006}), UK Infra-Red Telescope (UKIRT) Hemisphere Survey (UHS: \citealt{Dye 2018}) and Visible and Infrared Survey Telescope for Astronomy Hemisphere Survey (VHS: (\citealt{McMahon 2013})).
\begin{figure*}
\figurenum{2}
\includegraphics[scale=0.5,angle=0]{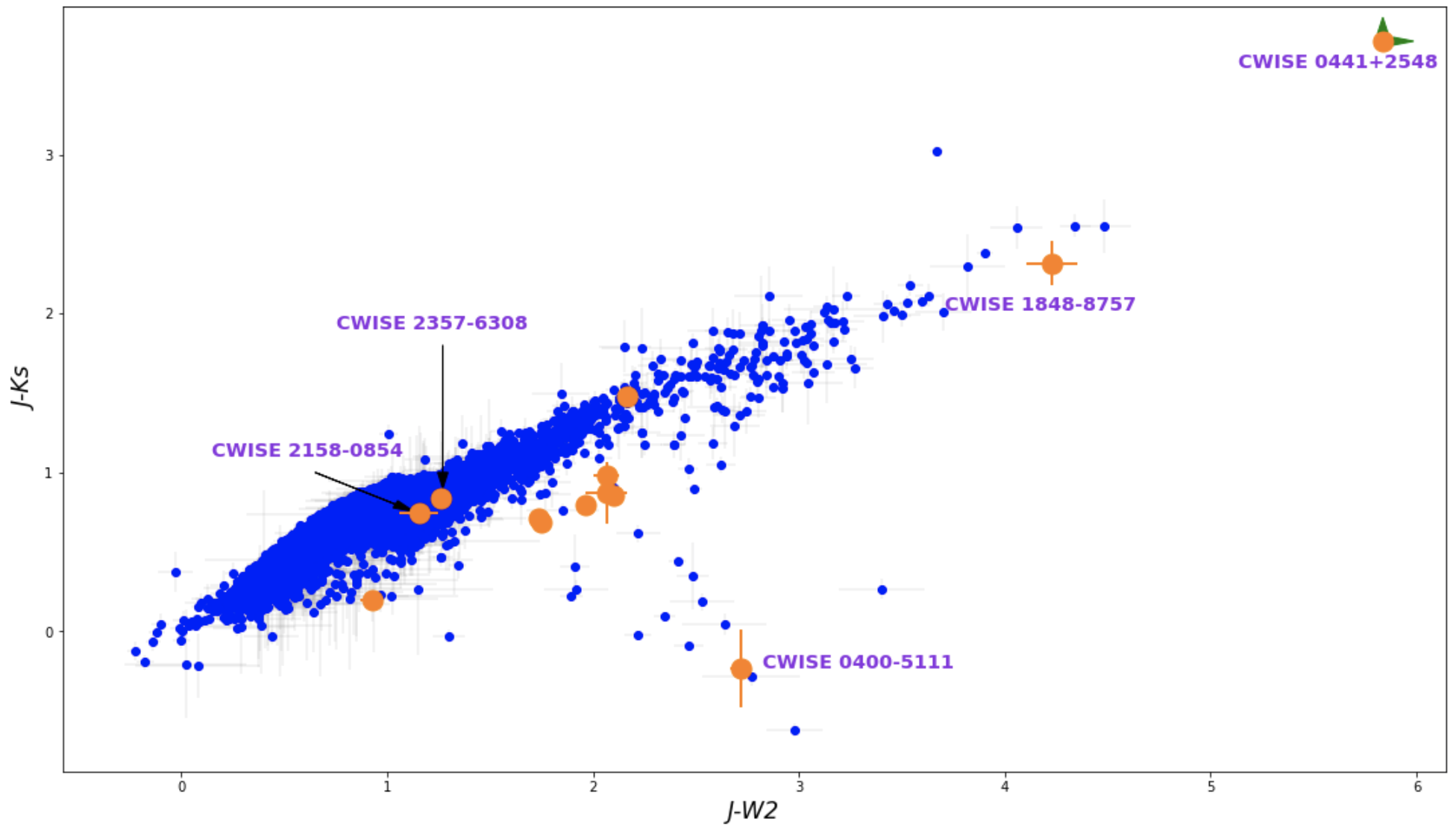}
\centering
\caption{J-W2 vs. J-Ks color-color diagram. The orange points are discoveries announced in this paper. For comparison, other proper motion discoveries from Figure 7 of \cite{Kirkpatrick 2016} are plotted as blue points. The green arrows denote color limits. Objects labeled in purple are discussed in the text. Note that the UHS discoveries are not plotted on this diagram because they have no Ks data.
\label{ColorColor1}}
\end{figure*}
\begin{figure*}
\figurenum{3}
\includegraphics[scale=0.5,angle=0]{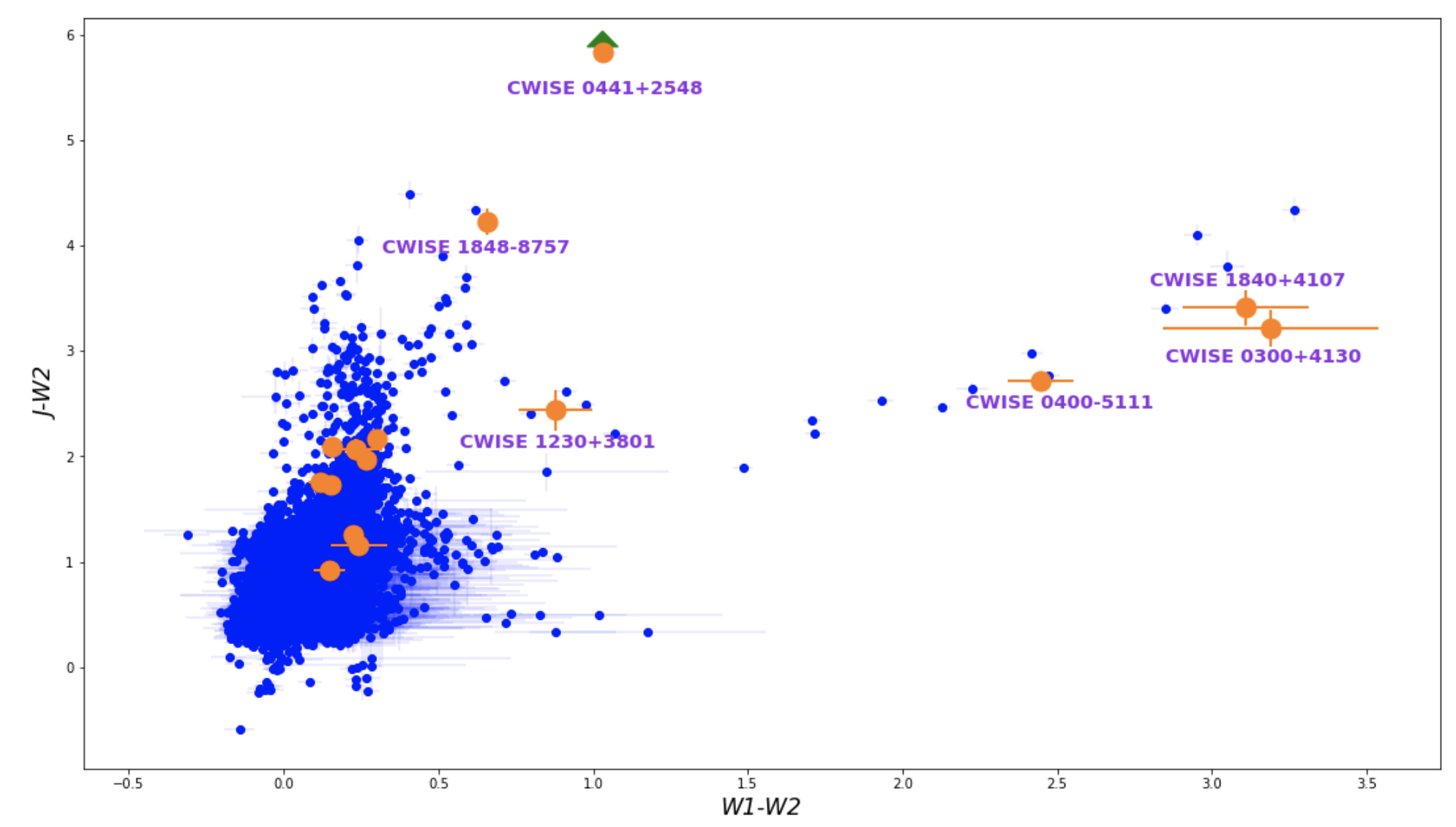}
\centering
\caption{W1-W2 vs. J-W2 color-color diagram. The orange points are discoveries announced in this paper. For comparison, other proper motion discoveries from Figure 8 of \cite{Kirkpatrick 2016} are plotted as blue points. The green arrows denote color limits. Objects labeled in purple are discussed in the text.
\label{ColorColor2}}
\end{figure*}
\par
Using the CatWISE2020 proper motion values, we propagated the position of each source in time to predict the position at the epoch of the database we were querying. Using a cone search of 5 arcseconds around the predicted position, we collected the J (1.25 $\mu$m), H (1.65 $\mu$m), and Ks (2.2 $\mu$m) magnitudes. (UHS only has J-band coverage.) We used multiple databases to maximize our chances of finding ancillary data, as not all data sets cover the entire sky or probe deeply enough to detect our objects. 
\par
After we matched our candidate list with the various infrared databases, we searched for mismatched candidates. Due to the difference in epochs and uncertainty in the proper motion measurements, our predicted location was not always accurate. 2MASS, with its large epoch difference from CatWISE2020 (15-18 yr), gave us the most spurious matches. 
 We used WiseView to verify or refute matches.
\subsection{Color-Color Diagrams}
\par
The reference color-color diagrams we used are Figures 7 and 8 in \cite{Kirkpatrick 2016}, which plot J-W2 vs.\ J-Ks and W1-W2 vs.\ J-W2 respectively. These plots contain all identified proper motion objects in the AllWISE1 (\citealt{Kirkpatrick 2014}) and AllWISE2  (\citealt{Kirkpatrick 2016}) motion surveys with magnitudes fainter than the nominal WISE W1 saturation limit of 8.1 mag. F, G, K, and M stars compose the dense locus while L and T dwarfs lie on the outer sections. For a complete description, please consult \cite{Kirkpatrick 2016}.
\par
Using TopCAT (\citealt{Taylor 2005}), we made a plot of our CatWISE2020 candidates with the same colors as the reference diagrams from \cite{Kirkpatrick 2016}. We compared our CatWISE2020 color-color plots side by side with the reference diagrams. As described further in section~\ref{Interesting objects}, we then extracted candidates that fell outside the loci of the F, G, K, or M stars determined in Figures 7 and 8 of \cite{Kirkpatrick 2016} and inputted them into a sublist. These are objects that have the highest probability of being cool brown dwarfs or other rare objects. 

\par
Next, to check the novelty of the objects recovered, we cross checked our entire sublist with SIMBAD (\citealt{Wenger 2000}) and other more recent or unpublished data sets (20 parsec census list from \citealt{Kirkpatrick 2021}, Backyard Worlds candidate list (\citealt{Kuchner 2017}), etc.) that have not yet been ingested in SIMBAD. Any previously discovered objects were removed from our sublist.
\par
Finally, we created finder charts of the new discoveries using a publicly available PYTHON program\footnote{Access Link: \href{https://github.com/jgagneastro/finder_charts}{Finder Charts} } These charts show the \textit{WISE} images in all four  bands and their counterparts (when there was coverage) in the three bands of 2MASS, three bands of VHS, three bands of the Digitized Sky Survey (DSS), 1 band of UHS, 5 bands of the Pan-Starrs (\citealt{Chambers 2016}) survey, and 5 bands of the Dark Energy Survey (DES; \citealt{Abbott 2018}). First, we used the charts to further verify or refute matches. Second, we used the charts to ensure our candidates had photometric properties across a wider wavelength range that were consistent with the classifications derived from their J-W2, J-Ks and W1-W2 colors.

Note that we only looked for interesting objects that had ancillary data at J, H, or Ks bands. Any object lacking UHS/VHS/2MASS data, or lacking a detection despite imaging in those data sets, will be part of a future study.

\section{Interesting objects found\label{Interesting objects}}
Our entire search yielded 16 new objects. 
All of the new discoveries are plotted on two color-color diagrams -- J-W2 vs. J-Ks (Figure \ref{ColorColor1}) and W1-W2 vs. J-W2 (Figure \ref{ColorColor2}). Outlying objects identified on these plots are discussed further below. The coordinates and photometry of our 16 discoveries are given in Table~\ref{other_discoveries}. Additionally, when available, Gaia photometry and parallaxes of our 16 discoveries are given in Table~\ref{Gaia_table}.

\subsection{CWISE J215859.63-085441.9 and CWISE J235713.17-630827.3}
\par
CWISEJ215859.63$-$085441.9 and CWISEJ235713.17$-$630827.3 are two nearby M dwarfs that are missing from our census of the solar neighborhood. After comparing the J-W2 colors of our objects to the data in Table 5 of \cite{Pecaut 2013}\footnote{See also: \url{http://www.pas.rochester.edu/~emamajek/EEM_dwarf_UBVIJHK_colors_Teff.txt}}, we find that their colors most closely match those of M4 and M5 dwarfs respectively.

Both objects were detected by Gaia EDR3 (\citealt{GAIA}) but had no reported parallaxes. This may suggest that these objects are a part of multiple systems, confounding the short time baseline astrometry currently available to Gaia EDR3. 

\begin{figure*}
\figurenum{4}
\includegraphics[scale=0.4,angle=0]{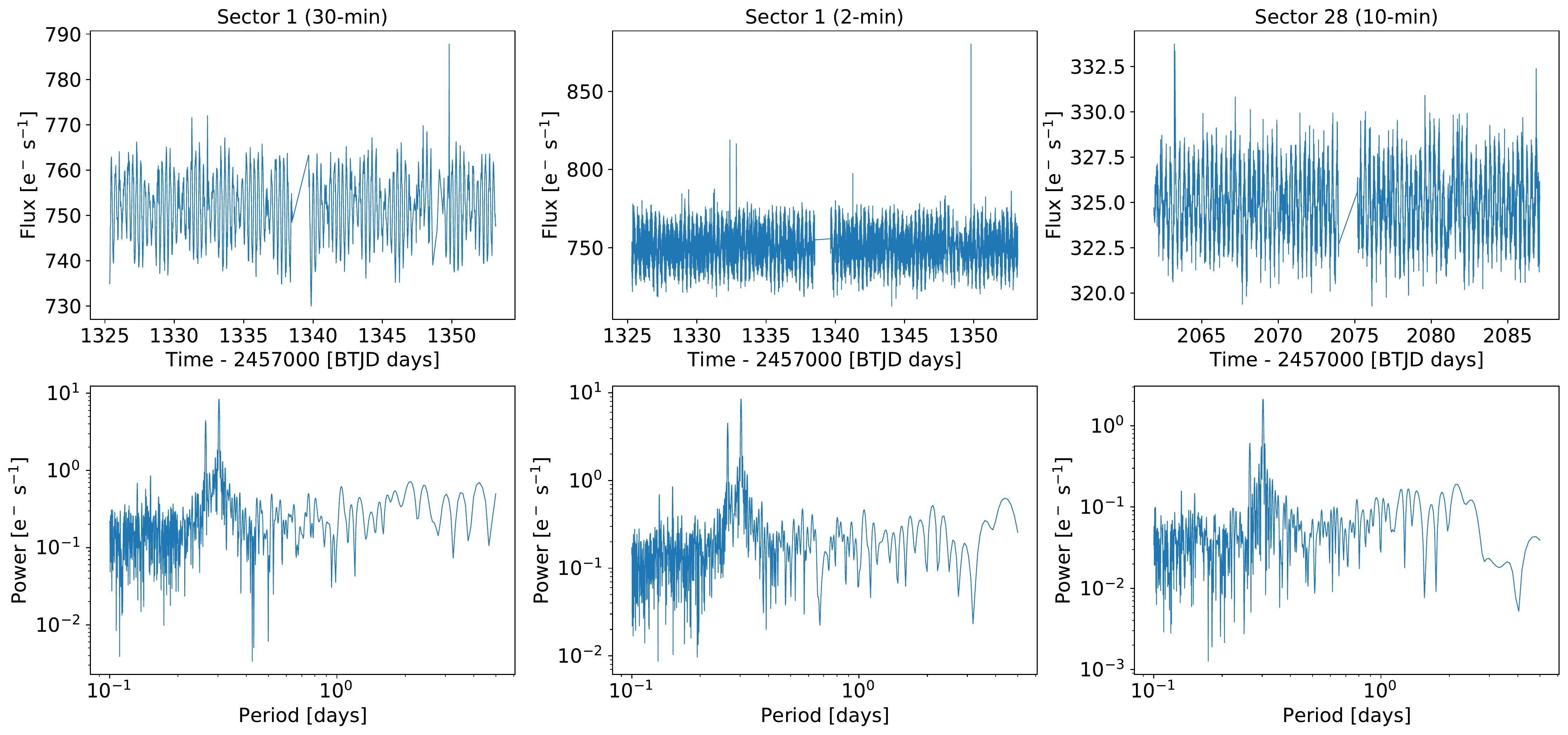}
\centering
\caption{Time series photometry of CWISE J235713.17$-$630827.3. The first row illustrates the entire light curve in three different sectors. The second row illustrates the periodogram for each of the light curves above.
\label{Tess Light Curve}}
\end{figure*}

We collected Transiting Exoplanet Survey Satellite (TESS; \citealt{Ricker 2015}) time series photometry on CWISE J235713.17$-$630827.3 to explore this hypothesis. We note that the location of CWISE J215859.63$-$085441.9 is currently not observed by TESS (at the time of this writing, September, 2021). The CWISE J235713.17$-$630827.3 light curve is shown in Figure \ref{Tess Light Curve}. The full light curve is shown in the top panel, and the Lomb-Scargle periodogram is shown in the bottom panel. The leftmost column is the Sector 1 30-min light curve. The middle column shows the same plots but for the Sector 1 2-min light curve. Finally, the rightmost column follows the same convention as the other two, but for the Sector 28 10-min light curve.
\par

There are two high powered peaks in each periodogram with periods of P1 = 0.3028 +/- 0.0004 d and P2 = 0.2646 +/- 0.0003 d. The implied periods are consistent over the two year baseline. The cause for the variation is likely to be star spots rotating into and out of view. There are two possible explanations for the double period. First, the object may be experiencing differential rotation like our Sun, leading to detect two distinct periods. Another explanation is the object is a system comprised of two stars with two slightly different rotation periods. Above we surmised that Gaia reported no parallax because the object may be a multiple system. The second explanation above for the TESS double period provides some credence to this statement. The objects seen by TESS are likely to be similar in magnitude because both variations are detected in a single light curve.

Since no Gaia parallax is available, we estimate the distance as follows. With a J-W2 vs. M$_J$ relation for M dwarfs
constructed using data from \cite{Pecaut 2013}, we estimate the distance of CWISE J235713.17-630827.3 if it is an equal magnitude double to be 27.5 parsecs. (Note, if single, the distance estimate would be $\sim$ 23 pc.) In addition, we found two sources in Gaia EDR3 which may correspond to the two components of our proposed physical pair. The separation between sources is $0{\farcs}24$. Using the G$_{BP}$- G$_{RP}$ vs M$_G$ relation found in Table 4 of \cite{Kiman 2019}, we estimate a M$_G$ value of 11.86 mag and spectral type of M5 for both Gaia sources. Then, by using the apparent G magnitude, we estimate distances of $\sim$33 and $\sim$37 pc for the two components of the binary.

Since CWISE J215859.63-085441.9 lacks a Gaia parallax, we theorize that it is also a binary like CWISE J235713.17-630827.3. Again, assuming an equal magnitude pair and using the J-W2 vs. M$_J$ relation for M dwarfs constructed using data from \cite{Pecaut 2013}, the distance estimate for CWISE J215859.63-085441.9 is $\sim$20.5 pc. (Note, if single, the distance estimate would be 14.5 pc.) We found only one source in Gaia EDR3. Using the G$_{BP}$- G$_{RP}$ vs. M$_G$ relation found in Table 4 of \cite{Kiman 2019}, we estimate a M$_G$ value of 12.92 mag and a spectral type of M6. Then, by using the apparent G magnitude we estimate a distance of $\sim$29 pc
\par
\begin{figure*}
\figurenum{5}
\plotone{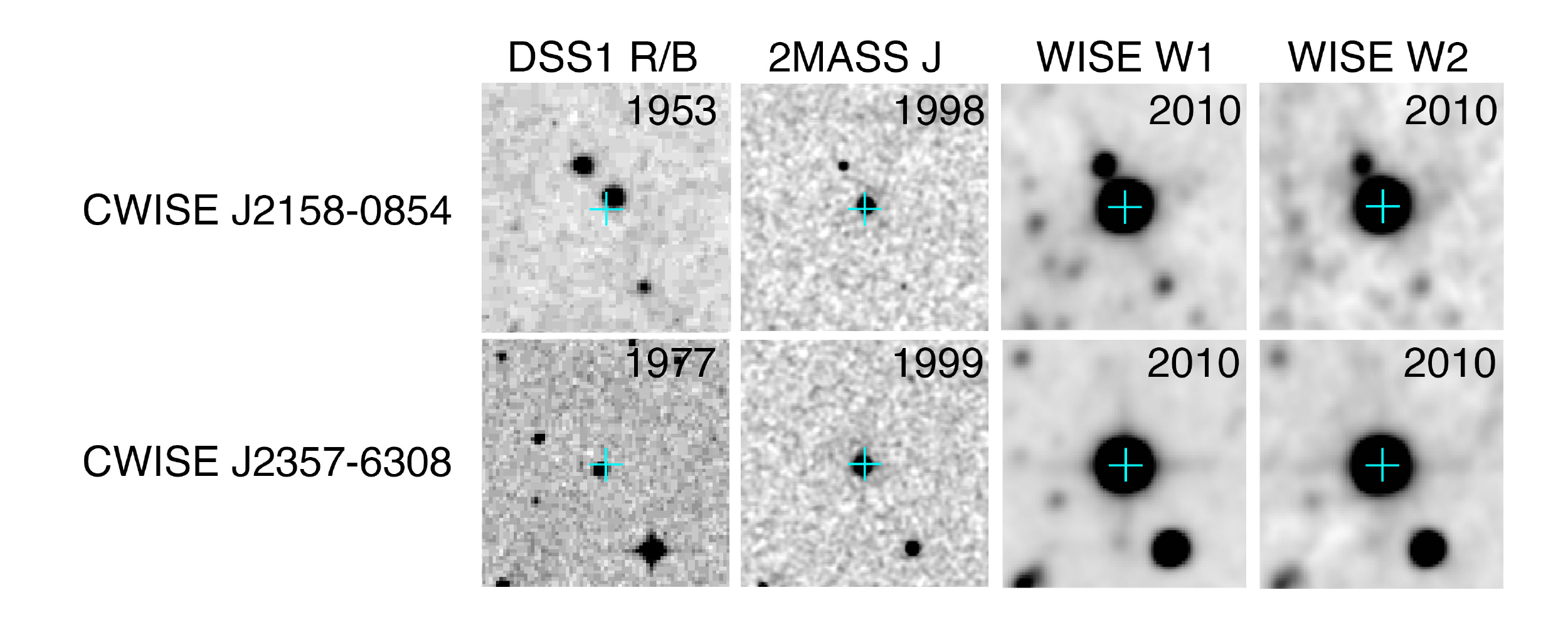}
\caption{Finder charts for CWISE J2158$-$0854 and CWISE J2357$-$6308. The finder chart shows a $>$30yr baseline of time. DSS2 R, 2MASS J, WISE W1, and WISE W2 images are shown. The objects have small but clear proper motions. Each cutout is 2 arcminutes on a side with north up and east to the left.}
\label{M dwarf Finder}
\end{figure*}

Despite their proximity to the Sun, these two M dwarfs were not discovered until now due to their small proper motions of $\mu_{tot} = 135{\pm}36$ mas/yr  for CWISE J215859.63$-$085441.9 and  $\mu_{tot} = 53{\pm}5$ mas/yr for CWISE J235713.17-630827.3, as measured by CatWISE. Finder charts of the two M dwarfs (Figure \ref{M dwarf Finder}) show the slow proper motion. Using the CatWISE2020 proper motion data and the distance estimate we calculated above, we estimated the tangential velocity of CWISE J215859.63$-$085441.9 to be 6.1 km/s. 
Again, assuming an equal magnitude pair, the tangential velocity estimate for CWISE J235713.17-630827.3 is 4.5 km/s. 

\par
The two M dwarfs were initially flagged as outliers when we plotted them using their reported VHS magnitudes. However, these magnitudes differ markedly from the 2MASS magnitudes. For example, CWISE J215859.63$-$085441.9 has VHS magnitudes of J$_{MKO}$=$11.868{\pm}0.001$ and Ks$_{VHS}$=$11.205{\pm}0.001$ mag, whereas, it has 2MASS magnitudes of J$_{2MASS}$=$11.215{\pm}0.026$ and Ks$_{2MASS}$=$10.470{\pm}0.025$ mag. The differences between the VHS and 2MASS magnitudes are $\Delta$J=0.653 and $\Delta$Ks=0.735 mag. This difference is larger than what we can attribute to the difference in filter sets. We believe the 2MASS data are credible and the VHS data problematic for the following reasons: (1) We know that 2MASS magnitudes are valid between $-4 < Ks< 16 $ mag, \footnote{See: \url{https://irsa.ipac.caltech.edu/data/2MASS/docs/releases/allsky/doc/sec2_2b.html}} and our source falls within this range. (2) VHS data has a bright limit (11.5 to 12 mag) where sources saturated and where the reported magnitudes cannot be properly measured (\citealt{Gonzales 2018}). Our objects are brighter than this limit. Thus, the difference in magnitudes is caused by saturation in the VHS data. Using the 2MASS magnitudes, the colors of both objects are consistent with M dwarfs and they are no longer color- color outliers.




\par 

\subsection {CWISE J123041.80+380140.9}
CWISE J123041.80+380140.9 is a high proper motion object ($\mu_{tot} = 764{\pm}115$ mas/yr). We used the W1-W2 color of {0.879$\pm$0.117} mag and the color vs. spectral type relation found in Table 13 of \cite{Kirkpatrick 2021} to estimate a spectral type of $\sim$T2. Based on Figure \ref{ColorColor2}, the $J-$W2 vs. W1-W2 location is consistent with an early T dwarf. 
\par
\begin{figure*}
\figurenum{6}
\plotone{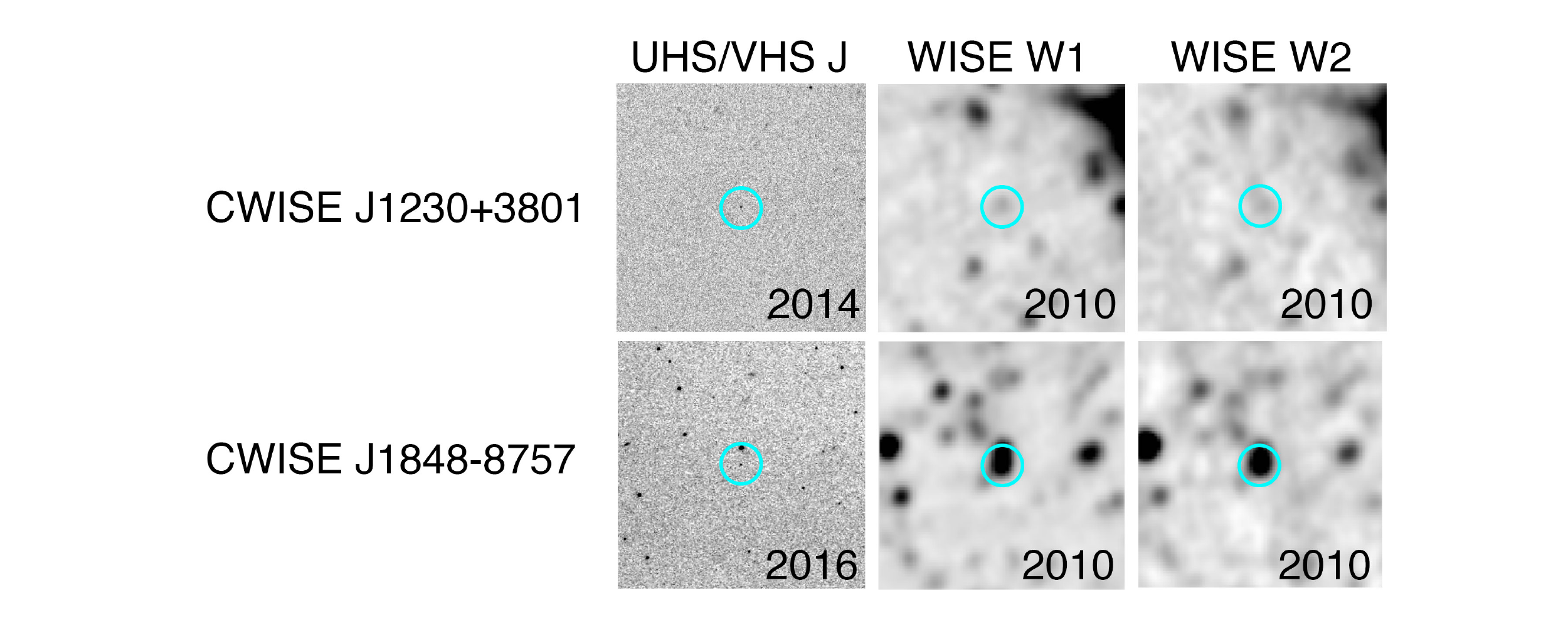}
\caption{Finder charts for CWISE J1230+3801 and CWISE J1848$-$8757. The finder charts illustrate the faintness of the objects. The UHS J, WISE W1, and WISE W2 images are shown for CWISE J1230+3801. The VHS J, WISE W1, and WISE W2 images are shown for CWISE J1848-8757. Each cutout is 2 arcminutes on a side with north up and east to the left.}
\label{T dwarf Finder}
\end{figure*}

The finder chart (Figure \ref{T dwarf Finder}) of our object illustrates its faintness. The object is not detected by 2MASS, but it is faintly detected in UHS and more clearly detected in WISE. 

\par
From the same paper and table, we used the absolute magnitude vs. color relation to derive an absolute W2 value of 12.3 mag, which implies a distance of $\sim$74 pc, using the apparent W2 magnitude. Using the CatWISE2020 proper motion and distance estimate, we calculate a tangential velocity of $\sim$234 km/s. We theorize that the large velocity of this object is likely caused by dynamical heating, which implies an old age. Furthermore, old age often implies low metallicity. Oddly, however, the object falls on the normal T dwarf track, so future spectroscopic observations are needed to reveal its nature.

\subsection{CWISE J184842.02-875747.9}
CWISE J184842.02-875747.9\footnote{CWISE J184901.34-875753.4 lies 11.2 arcseconds to the southeast. Despite the CatWISE2020 catalog measuring no significant proper proper motion, it appears to be moving in the WiseView blink. However this proper motion is not aligned with CWISE J184842.02-875747.9.} falls along the L dwarf track as seen in Figure \ref{ColorColor1} and Figure \ref{ColorColor2}. A finder chart is shown in Figure \ref{T dwarf Finder}. We used the W1-W2 color of {0.659$\pm$0.026} mag and the color vs. spectral type relation found in Table 13 of \cite{Kirkpatrick 2021} to estimate a spectral classification of $\sim$L7.5. From the same paper and table, we used the absolute magnitude vs. color relation to derive an absolute W2 value of 12.2 mag, which implies a distance of $\sim$29 pc, using the apparent W2 magnitude. A color of J-W2=$4.23{\pm}0.13$ mag is surprisingly red for an L dwarf. The only L dwarfs that are this red on Figure 14 of \cite{Kirkpatrick 2021} are young objects. In addition, the only L dwarfs that rival this J-W2 color in Figure 8 of \cite{Faherty 2016} are young objects. Using the W2$_{YNG}$ vs. Spectral Type relation found in Table 19 of \cite{Faherty 2016} we derived a absolute W2 value of 10.8 mag, which implies a distance of$\sim$ 54 pc, using the apparent magnitude. If the object is determined to be young, the true distance should be closer to this value than the 29 pc distance value estimated above. Using the CatWISE2020 proper motion ($\mu_{tot} = 158{\pm}15$ mas/yr) and the distance estimate if the object is young, we calculate a tangential velocity of $\sim$35 km/s which would not run counter to the young classification. We therefore propose this is likely a candidate young L dwarf. However, using the CatWISE 2020 proper motion data, we find no clear membership in any young nearby moving group from Banyan $\Sigma$ (\citealt{Gagne 2018}), and it is determined to be a field object with 99 \% probability. Future spectroscopic observations of CWISE J184842.02-875747.9 would help determine whether or not it is young via low-gravity spectral features (\citealt{Allers 2013}).
\subsection{Additional objects discovered\label{Additional objects}}

The rest of the 16 objects are discussed further below:


\begin{figure*}
\figurenum{7}
\plotone{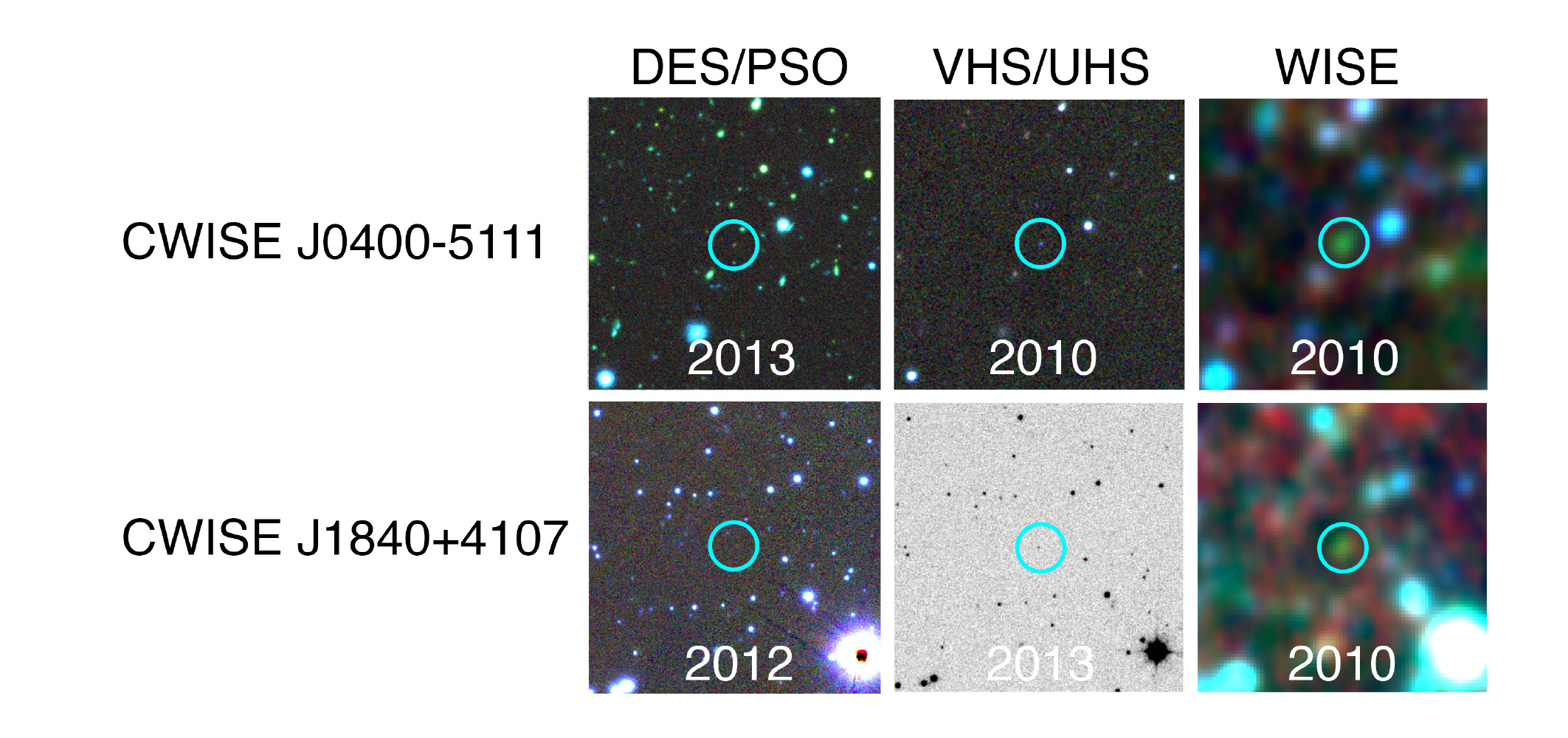}
\caption{Finder charts for CWISE J0400$-$5111 and CWISE J1840+4107. For CWISE J0400-5111 these are three-color images made from DES g/i/y, VHS J/H/Ks and WISE W1/W2/W3. For WISE J1840$+$4107 the images are PSO g/i/y, UHS J and WISE W1/W2/W3. (Only one band was available for UHS.) For CWISE J0400-5111, the red color in the optical (DES), the blue color in the near infrared (VHS) and the green color in the mid infrared (WISE) parallel the spectral energy distribution of a T dwarf. CWISE J1840+4107 has less information than CWISE J0400$-$5111 but is faint in the optical, detected at J, and brighter and green in the WISE bands, again characteristic of a T dwarf.
\label{T dwarf Finder 2}}
\end{figure*}

\begin{itemize}
    \item CWISE J044109.35+254854.4 falls in an unusual spot in Figure~\ref{ColorColor1} and Figure~\ref{ColorColor2} at colors of J-W2>5.8 mag, J-K>3.7 mag and W1-W2=$1.031{\pm}0.016$ mag. Objects that live in this portion of the diagram are either extremely red L dwarfs or reddened objects (cf. figure 7 and 8 of \citealt{Kirkpatrick 2016}).  The coordinates place the object in a region of nebulosity toward the Taurus Molecular Cloud. Thus, we surmise the object is behind the cloud and is being reddened by it. CWISE J044109.35+254854.4 has a motion significance of 21 $\sigma$ ($\mu_{tot} = 160{\pm}7.5$ mas/yr). In addition, this source appears point-like and exhibits clear motion between the 1998-epoch 2MASS Ks image and the 2010-epoch AllWISE W1/W2 images, further confirming the proper motion measured by CatWISE2020. A spectrum of this object is needed to measure a spectral type, measure the amount of extinction, estimate a distance, and confirm if the distance is sufficient to place the object behind the cloud.
    \item CWISE J030009.79+413051.2 is likely a late T dwarf. This object was not detected by 2MASS and was only found in UHS and WISE. We used the J-W2 color of {3.22$\pm$0.18 mag} and the color vs. spectral type relation found in Table 13 of \cite{Kirkpatrick 2021} to estimate a spectral classification of T8. As shown in Figure \ref{ColorColor2}, this object falls along the normal late-T dwarf track. From the same paper and table, we used the absolute magnitude vs. color relation to derive an absolute W2 value of 13.0 mag, which implies a distance of $\sim$40 pc, using the apparent W2 magnitude.
    \item
    CWISE J040052.80-511142.5 and CWISE J184048.17+410727.4 have a motion significance of only 1.3$\sigma$ and 1.0 $\sigma$ respectively in CatWISE 2020 (see Table~\ref{other_discoveries}) giving us little confidence that the objects are nearby. However, when querying NOIRLab Source Catalog, we find that CWISE J040052.80-511142.5 has a motion significance of 4.3 $\sigma$ ($\mu_{tot} = 98{\pm}23$ mas/yr). The location of CWISE J184048.17+410727.4 is unfortunately not covered by the NOIRLab Source Catalog. However in Figure \ref{T dwarf Finder 2} we see that the overall colors from the optical to the mid infrared of both objects correspond to T dwarfs. Using the methodology above, we estimated spectral types of T7.5 for CWISE J040052.80-511142.5 and T8.5 for CWISE J184048.17+410727.4, which correspond to 31 and 27 pc, respectively. To confirm these classifications, spectra are needed.
    \item
    CWISE J103716.55-750257.7 and CWISE J125924.86-552531.8 are two likely L dwarfs.   We used the W1-W2 color of {0.23$\pm$0.07} mag and the color vs. spectral type relation found in Table 13 of \cite{Kirkpatrick 2021} to estimate a spectral classification of L2.5 for CWISE J103716.55-750257.7. From the same paper and table, we used the absolute magnitude vs. color relation to derive an absolute W2 value of 12.0 mag, which implies a distance of $\sim$37 pc, using the apparent W2 magnitude. Using the methodology above, CWISE J125924.86-552531.8 has a spectral type of L4 and a distance of $\sim$36 pc.
    \item CWISE J141545.74-330545.0 and CWISE J210159.84-783846.8 have measurements in all three colors found in Figure~\ref{ColorColor1} and Figure~\ref{ColorColor2}. Their locations in these figures when compared to Figure 7 and 8 of \cite{Kirkpatrick 2016} are consistent with L subdwarfs or early T dwarfs. Spectra are needed for both objects to determine their nature and to estimate distances
    \item CWISE J120657.58-311221.1 also has measurements in all three colors found in Figure~\ref{ColorColor1} and Figure~\ref{ColorColor2}. Its location is consistent with an M subdwarf. Using its Gaia EDR3 parallax, we calculate a distance of $\sim$156 parsecs. Then, using the Gaia EDR3 proper motion data (($\mu_{tot} = 534{\pm}23$ mas/yr) and distance, we calculate a very high tangential velocity of $\sim$394 km/s.
    \item  All of the following objects are outliers, but not for astrophysical reasons. CWISE J144053.88-121712.3 lies near another object leading to contaminated photometry. CWISE J052703.28-113056.7 and CWISE J015330.24-520046.6 are close comoving doubles. WISE has combined photometry from both sources while VHS reports photometry from both, which leads to inaccurate J-W2 colors.

\end{itemize}

\par


\section{Conclusion\label{Conclusion}}
\par We have presented 16 new dwarfs identified through supervised learning methods. Specifically, we highlight four interesting objects: 2 nearby M dwarfs (CWISE J215859.63$-$085441.9, CWISE J235713.17$-$630827.3), a high proper motion early T dwarf (CWISE J123041.80+380140.9) and likely a young L dwarf (CWISE J184842.02-875747.9).
\par
 Spectroscopy will help to reveal the nature of these objects. For the M dwarfs, this will be used to get more definitive spectral classifications and distance estimates. We also await future Gaia data to confirm the binary hypothesis. For the high proper motion T dwarf, a spectrum would confirm the spectral type and metallicity. Finally, a spectrum of the suspected young L dwarf would be used to search for features of low gravity, which is a direct indicator of youth.

There are a few limitations to our search method described in \S~\ref{Candidate Selection}. First, given we are searching for nearby objects through their significant proper motions, we are not sensitive to nearby objects with small, insignificant proper motions.
Second, some of our objects are not detected by 2MASS. Of these, some lack deeper UHS or VHS data, while others are undetected in their UHS or VHS images. Without detections in these data, we are not able to use our color-color diagrams to pick out interesting objects. Additional analysis must also be done on the objects in the candidate list with UHS and VHS imaging but no detection in their respective images. This means that UHS and VHS failed to detect the objects at the sensitivities of those surveys. This is relevant because the elusive Y dwarfs (\citealt{Kirkpatrick 2012}) may be tucked away in this exclusive list of objects. 

This paper represents just one way to search through the CatWISE2020 data. The entire CatWISE2020 and Backyard Worlds team is employing a myriad of methods for analyzing WISE data and this will culminate in new insights about the luminosity and mass functions (Kirkpatrick et al. in prep). Our approach is just one machine learning method applied to WISE \textit{catalog} data. Other machine learning methods have used different training sets to explore new areas of parameter space or they have used WISE \textit{imaging} data (Caselden in prep). There are other approaches that do not use machine learning such as using catalog selection (\citealt{Meisner 2020}) directly from IRSA or human vetting of motion candidates (\citealt{Kuchner 2017}). All of these methods will help researchers fully explore the almost 1.8 billion sources in the CatWISE2020 catalog and the epochal coadds from the unWISE collection, and undoubtedly aid in uncovering even more interesting objects.
\par Our research illustrates that, despite our best efforts over the last decade, there are still discoveries near the Sun to be made. In addition, our census of the faintest stellar and sub-stellar objects is still incomplete. To finish the census, astronomers (and the public!) are encouraged to continue exploring CatWISE2020 and exploit its capabilities.

\section{acknowledgements}

CatWISE is funded by NASA under Proposal No. 16-ADAP16-0077 issued through the Astrophysics Data Analysis Program, and uses data from the NASA-funded \textit{WISE} and NEOWISE projects.  This research has made use of the NASA/IPAC Infrared Science Archive, which is funded by the National Aeronautics and Space Administration and operated by the California Institute of Technology. The Backyard Worlds: Planet 9 team would like to thank the many Zooniverse volunteers who have participated in this project, from providing feedback during the beta review stage to classifying flipbooks to contributing to the discussions on TALK. We would also like to thank the Zooniverse web development team for their work creating and maintaining the Zooniverse platform and the Project Builder tools. F.M. acknowledges support from grant 80NSSC20K0452 under the NASA Astrophysics Data Analysis Program. This research was supported by NASA grant 2017-ADAP17-0067. This material is based upon work supported by the National Science Foundation under Grant No. 2007068, 2009136, and 2009177. This research has made use of the SIMBAD database, operated at CDS, Strasbourg, France. This publication makes use of data products from the Two Micron All Sky Survey, which is a joint project of the University of Massachusetts and the Infrared Processing and Analysis Center/California Institute of Technology, funded by the National Aeronautics and Space Administration and the National Science Foundation. This publication makes use of data products from the Near-Earth Object Wide-field Infrared Survey Explorer (NEOWISE), which is a joint project of the Jet Propulsion Laboratory/California Institute of Technology and the University of Arizona. NEOWISE is funded by the National Aeronautics and Space Administration. Our research used data from the UHS collaboration and the VISTA Hemisphere Survey (VHS).
We want to thank the Student Astrophysics Society\footnote{\url{https://www.studentastrophysicssociety.com}} for providing the resources that enabled the pairing of high school and undergraduate students with practicing astronomers and advanced citizen scientists. We thank the anonymous referee for insightful suggestions that helped improve the manuscript.  
\facilities{WISE/NEOWISE, 2MASS, UKIRT, VISTA}
\software{Astropy: \cite{Astropy 2013}, Banyan $\Sigma$: \cite{Gagne 2018}, TOPCAT: \cite{Taylor 2005} XGBoost: \cite{Chen 2016}, WiseView: \cite{Caselden 2018}}



\begin{longrotatetable}
\centering
\begin{deluxetable*}{lcccclcccccccc}
\tabletypesize{\scriptsize}
\tablecaption{Data for New Discoveries\label{other_discoveries}}
\tabletypesize{\tiny}
\tablehead{
\colhead{CatWISE Designation} &
\colhead{J2000 RA} &
\colhead{J2000 Dec} &
\colhead{$\mu_\alpha$}&
\colhead{$\mu_\delta$}&
\colhead{W1\tablenotemark{a}}&
\colhead{W2\tablenotemark{a}}&
\colhead{$J$}&
\colhead{$H$}&
\colhead{$K$}&
\colhead{$JHK$ Ref.} &
\colhead{Classification} &
\colhead{BYW Codiscovers\tablenotemark{c}}
\\
\colhead{} &
\colhead{(deg)} &
\colhead{(deg)} &
\colhead{(asec yr$^{-1}$)} &
\colhead{(asec yr$^{-1}$)} &
\colhead{(mag)} &
\colhead{(mag)} &
\colhead{(mag)} &
\colhead{(mag)} &
\colhead{(mag)} &
\colhead{} &
\colhead{} &
\colhead{}
\\
\colhead{(1)} &       
\colhead{(2)} &
\colhead{(3)} &
\colhead{(4)} &
\colhead{(5)} &
\colhead{(6)}&
\colhead{(7)}&
\colhead{(8)}&
\colhead{(9)} &
\colhead{(10)} &
\colhead{(11)} &
\colhead{(12)} & 
\colhead{(13)}
}
\startdata
CWISE J015330.24-520046.6 & 28.3760178 & $-$52.0129465 & $0.12459{\pm}0.0075$ &	$0.04784{\pm}0.0065$ & $14.202{\pm} 0.013$ & $14.050{\pm} 0.013$ & $15.783{\pm}	0.005$ &	\nodata &	$15.076{\pm}	0.009$ & VHS & \tablenotemark{b}\\
CWISE J030009.79+413051.2 & 45.0408097 & 41.5142428 & $-0.26243{\pm} 0.0857$& $0.04321{\pm} 0.0838$& $19.173{\pm} 0.343$& $15.984{\pm} 0.065$& $19.199{\pm}0.165$& \nodata& \nodata & UHS\tablenotemark{d} & T dwarf\\
CWISE J040052.80-511142.5 & 60.2200208 & $-$51.1951632 & $0.03525{\pm} 0.0435$	& $-0.06973{\pm}0.0443$  &  $18.274{\pm} .098$ & $15.828 {\pm} 0.037$ &$ 18.547{\pm}	0.036$ & $18.863{\pm} 0.232$ &	$18.778{\pm}	0.243$ & VHS & T dwarf & B\\
CWISE J044109.35+254854.4& 70.2889956 & 25.8151209 & $0.14188{\pm} 0.0053$ & $-0.08250{\pm}0.0053$ & $13.181 {\pm} 0.013$ & $12.150 {\pm} 0.010$& $>$17.987 & $16.324{\pm} 0.022$ & $14.274 {\pm} 0.074$ & 2MASS & reddened star\\
CWISE J052703.28-113056.7 & 81.7636813 & $-$11.5157684 & $0.01655{\pm} 0.0185$&	$-0.23600{\pm} 0.0188$ & $14.687 {\pm}0.027$ & $14.569 {\pm}0.027$ & $16.320{\pm}	0.011$ &	\nodata &	$15.636{\pm}	0.024$ &VHS & \tablenotemark{b}\\
CWISE J103716.55-750257.7& 159.3189815 & $-$75.0493751 & $-0.19359 {\pm}0.0115$ & $0.02816{\pm} 0.0119$ & $15.118{\pm} 0.063$ & $14.884{\pm} 0.056$& $16.948{\pm}	0.020$ &	\nodata &	$15.967{\pm}0.041$ &VHS & L dwarf\\
CWISE J120657.58-311221.1& 181.7399330 & $-$31.2058758 & $-0.40234 {\pm}0.0204$ & $0.40390 {\pm}0.0233$ & $15.766 {\pm}0.022$ & $15.616{\pm} 0.046$ & $16.542{\pm}	0.027$ &	\nodata&	$16.343{\pm}	0.055$ & VHS & M subdwarf\\
CWISE J123041.80+380140.9 & 187.6741736 & 38.0280398 & $0.14901{\pm}0.0743$ & $-0.74981{\pm}0.0880$ & $17.559 {\pm}0.069$& $16.680 {\pm}0.095$&  $19.117{\pm}	0.163$ & \nodata & \nodata & UHS & high proper motion T2 dwarf\\
CWISE J125924.86$-$552531.8 &194.8535993 & $-$55.4255028 & $-0.21910{\pm} 0.0126$	& $0.00887 {\pm}0.0144$& $15.126{\pm} 0.017$ & $14.826 {\pm}0.023$& $16.992	{\pm}0.022$ &	\nodata &	$15.515{\pm}	0.022$&VHS & L dwarf\\
CWISE J141545.74-330545.0& 213.9405948 & $-$33.0958410 &  $-0.17852{\pm} 0.0115
$	&$-0.21219{\pm} 0.0133$ & $14.941{\pm} 0.017$ &$14.672{\pm} 0.025$& $16.634{\pm}	0.018$	& \nodata &	$15.839{\pm}	0.032$ &VHS & L subdwarf or early T dwarf, A \\
CWISE J144053.88-121712.3 & 220.2245053 & $-$12.2867517 &  $-0.09991 {\pm}0.0136$	& $-0.23436 {\pm}0.0157$ & $15.063{\pm} 0.019$ & $14.906{\pm} 0.030$ & $17.004{\pm}	0.0165$ &	$16.443{\pm}	0.029$ &	$16.146{\pm}	0.036 $ & VHS & \tablenotemark{b}\\
CWISE J184048.17+410727.4 & 280.2007202 & 41.1242809 & $-0.00784{\pm} 0.0438 $&	$-0.06299{\pm}0.0479$ & $18.938{\pm} 0.199$& $15.829{\pm} 0.036$ & $19.243{\pm}	0.164$ & \nodata & \nodata& UHS & T dwarf\\\
CWISE J184842.02-875747.9 & 282.1750944 & $-$87.9633209 & $-0.01739 {\pm}0.0106$&	$-0.15779{\pm} 0.0107$ & $15.124 {\pm}0.019$ & $14.465 {\pm}0.017$& $18.699 {\pm}	0.124$& \nodata&	$16.382{\pm}	0.059$ &VHS & young L dwarf?& A,C\\
CWISE J210159.84-783846.8 & 315.4993413 & $-$78.6463487 & $0.27545 {\pm}0.0317$&$-0.18208 {\pm}0.0301$ & $16.489{\pm} 0.031$ & $16.254{\pm} 0.063$& $18.324{\pm}	0.079$ &	\nodata&	$17.451{\pm}	0.176$ &VHS & L subdwarf or early T dwarf\\
CWISE J215859.63$-$085441.9 & 329.7484963 & $-$8.9116413 & $0.0748{\pm}0.0044$ & $-0.11231{\pm}0.0040$&$10.303{\pm}0.013$&$10.060{\pm}0.089$&$11.215{\pm}0.026$&$10.708{\pm}0.023$&$10.470{\pm}0.025$ & 2MASS & nearby M dwarf\\
CWISE J235713.17$-$630827.3 & 359.3048900& $-$63.1409326&$-0.03229{\pm}0.0038$&$0.04222{\pm}0.0034$&$9.798{\pm}0.012$&$9.573{\pm}0.007$ & $10.813{\pm}0.023$ & $10.272{\pm}0.022$ & $9.997{\pm}0.023$ & 2MASS & nearby M dwarf\\
\enddata
\tablenotetext{a}{These are the W1 and W2 magnitudes measured while taking into account proper motion from the CatWISE2020 Catalog(listed as w1mpro\_pm and w2mpro\_pm)}
\tablenotetext{b}{Cannot be classified. See Section~\ref{Additional objects}}
\tablenotetext{c}{Reference code for discover: A = Caselden, B = Bickle and C = Goodman. These discoverers independently discovered objects in this paper using tools provided
and developed for the Backyard Worlds citizen science project.}
\tablenotetext{d}{Note that UHS is the only UKIRT survey our targets fell into.}
\end{deluxetable*}

\end{longrotatetable}


\begin{deluxetable*}{lcccclc}
\tabletypesize{\scriptsize}
\tablecaption{Gaia EDR3 data on new discoveries\label{Gaia_table}}
\tabletypesize{\tiny}
\tablehead{
\colhead{CatWISE Designation} &
\colhead{Gaia RA} &
\colhead{Gaia Dec} &
\colhead{$G$}&
\colhead{$G_{BP}$}&
\colhead{$G_{RP}$}&
\colhead{Plx}
\\
\colhead{} &
\colhead{(deg)} &
\colhead{(deg)} &
\colhead{(mag)} &
\colhead{(mag)} &
\colhead{(mag)} &
\colhead{(mas)} 
\\
\colhead{(1)} &       
\colhead{(2)} &
\colhead{(3)} &
\colhead{(4)} &
\colhead{(5)} &
\colhead{(6)}&
\colhead{(7)}
}
\startdata
CWISE J015330.24-520046.6 & 28.375773 & $-$52.012694 & $19.037842{\pm}0.003477$ &	$20.801437{\pm}0.166603$ & $17.803684{\pm}0.014892$ & $4.7697{\pm} 0.1993$\\
CWISE J052703.28-113056.7 & 81.763843 & $-$11.516011 & $19.325026{\pm} 0.003703$&	$20.786207{\pm} 	0.072447$ & $18.060532 {\pm}0.015927$ & $3.2509	 {\pm}0.2729$\\
CWISE J120657.58-311221.1& 181.739803 & $-$31.205797 & $19.547705 {\pm}0.003853$ & $20.997263 {\pm}0.114532$ & $18.291862 {\pm}0.020912$ & $	6.4006{\pm} 0.3416$\\
CWISE J215859.63$-$085441.9 & 329.748553 & $-$8.911683 & $14.468976{\pm}0.015447$ & $15.793684	{\pm}0.003651$&$12.920640{\pm}0.003923$ &\nodata\\
CWISE J235713.17$-$630827.3 \tablenotemark{a} & 359.304944& $-$63.140875&$14.482848{\pm}0.020162$&$16.019608{\pm}0.006164$&$12.760140{\pm}0.004215$& \nodata\\
CWISE J235713.17$-$630827.3 \tablenotemark{a} & 359.304881& $-$63.140937&$14.717904{\pm}0.004546$&$16.005678{\pm}0.007803$&$12.754520{\pm}0.004263$ & \nodata\
\enddata
\tablenotetext{a}{This CatWISE2020 object is split into two components in Gaia EDR3}
\end{deluxetable*}


\end{document}